\documentclass[conference]{IEEEtran}
\ifCLASSINFOpdf
\else
\fi
\usepackage{footnote}
\usepackage[pdftex]{graphicx}
\usepackage[square,comma,sort&compress,numbers]{natbib}
\usepackage{amsmath}
\usepackage{amssymb}
\IEEEaftertitletext{\vspace{-2\baselineskip}}
\begin{document}
\title{Multiple Access Analog Fountain Codes}
\author{\IEEEauthorblockN{Mahyar Shirvanimoghaddam, Yonghui Li and Branka Vucetic}
\IEEEauthorblockA{School of Electrical and Information
Engineering, The University of Sydney, NSW, Australia\\
Emails:\{mahyar.shirvanimoghaddam, yonghui.li, branka.vucetic\}@sydney.edu.au}}

\maketitle
\begin{abstract}
In this paper, we propose a novel rateless multiple access scheme based on the recently proposed capacity-approaching analog fountain code (AFC). We show that the multiple access process will create an equivalent analog fountain code, referred to as the multiple access analog fountain code (MA-AFC), at the destination. Thus, the standard belief propagation (BP) decoder can be effectively used to jointly decode all the users. We further analyze the asymptotic performance of the BP decoder by using a density evolution approach and show that the average log-likelihood ratio (LLR) of each user's information symbol is proportional to its transmit signal to noise ratio (SNR), when all the users utilize the same AFC code. Simulation results show that the proposed scheme can approach the sum-rate capacity of the Gaussian multiple access channel in a wide range of signal to noise ratios.
\end{abstract}
\IEEEpeerreviewmaketitle

\section{Introduction}
Multiple access technologies, enabling multiple users to share common radio resources to fulfill their own transmission requirements, have been widely studied over decades. Successive decoding has been proved to be efficiently used to approach the corner points of the well known capacity region of the multiple access channel (MAC), while the middle part is achieved by additional rate splitting or time sharing \cite{AchGausSpatCoup}. Moreover, multiuser communication techniques such as code-division multiple access (CDMA), that allow for more robust and less complex joint detection/decoding, have had limited success in achieving the inner points of the MAC capacity region. The problem will be more challenging when all the users transmit with the same rate and power, as there is no structural irregularity to initiate the decoding convergence due to the fact that all the users are operating under the same conditions .

Recently, rateless codes have been applied in the multiple access scenario, where different users use them to adapt to the time varying wireless channel. In existing rateless multiple access (RMA) schemes \cite{RatelessMAErasure,RMA2,RMA}, each user encodes its packets using a separate rateless code. Thanks to the rate adaptation property of rateless codes, without knowledge of channel state information (CSI) or frequent feedback information from the destination, RMA achieves an excellent performance on multiuser detection over both erasure and noisy channels. However, RMA conducts separate decoding for each user at the destination and soft decoding information is exchanged between several decoders in each iteration of the decoding algorithm; thus, not practical for systems with a large number of users.

Spatial coupling effects have been recently explored to increase the transmission efficiency in multiple access channels \cite{Mod-Sum,SigSag,NCRMA}. In \cite{SRAGraph}, spatial coupling was applied to the erasure random access channel and the iterative interference cancelation process has been modeled as the erasure recovery process of low-density parity-check (LDPC) codes over a binary erasure channel (BEC).  Although this approach can achieve a high throughput over erasure MACs, it cannot be used in wireless channels, where the collided packets are affected by fading and noise. To overcome this problem, authors in \cite{AchGausSpatCoup} showed that the capacity of a Gaussian MAC can be achieved by using a spatially coupled sparse graph multi-user modulation with an iterative interference cancelation scheme at the receiver. However, this approach performs decoding and interference cancelation separately and its complexity increases significantly as the number of active users increases.

In this paper, we propose a novel multiple access scheme based on a newly proposed capacity approaching analog fountain code (AFC) \cite{MahyarLetter} to approach the sum-rate capacity of the multiple access. In the proposed scheme, referred to as multiple access analog fountain code (MA-AFC), each user uses an AFC code to generate a potentially limitless number of coded symbols. As the sum of coded symbols from various users is received at the destination, the received signals at the destination can be seen as coded symbols of an equivalent AFC code with a larger code degree. As a result, the standard belief propagation (BP) decoder can be used to jointly decode all the users at the destination.  We analyze the decoding performance through density evolution analysis and show that the average LLR of each user is proportional to its transmit SNR. Simulation results show that the proposed scheme can approach the sum-rate capacity of the Gaussian multiple access channel in a wide range of SNRs and can be effectively applied to systems with a large number of users due to the linear complexity of the encoding and decoding of AFCs. Throughout the paper, we use boldface letters to denote the vectors, where the $i^{th}$ entry of vector \textbf{v} is denoted by $v_i$. Matrices are represented by boldface capital letters, $\textbf{G}$, where $g_{i,j}$ is the $j^{th}$ entry in the $i^{th}$ row of $\textbf{G}$ and $\textbf{G}^{'}$ is the transpose of \textbf{G}.

The rest of the paper is organized as follows. In Section II, we present the system model. The proposed multiple access scheme is presented in Section III. Section IV presents the decoding algorithm and the asymptotic analysis of the belief propagation decoding of the MA-AFC scheme based on the density evolution analysis. Simulation results are shown in Section V, and finally conclusions are drawn in Section VI.

\section{System Model}
\begin{figure}[!t]
\centering
\includegraphics[scale=0.32]{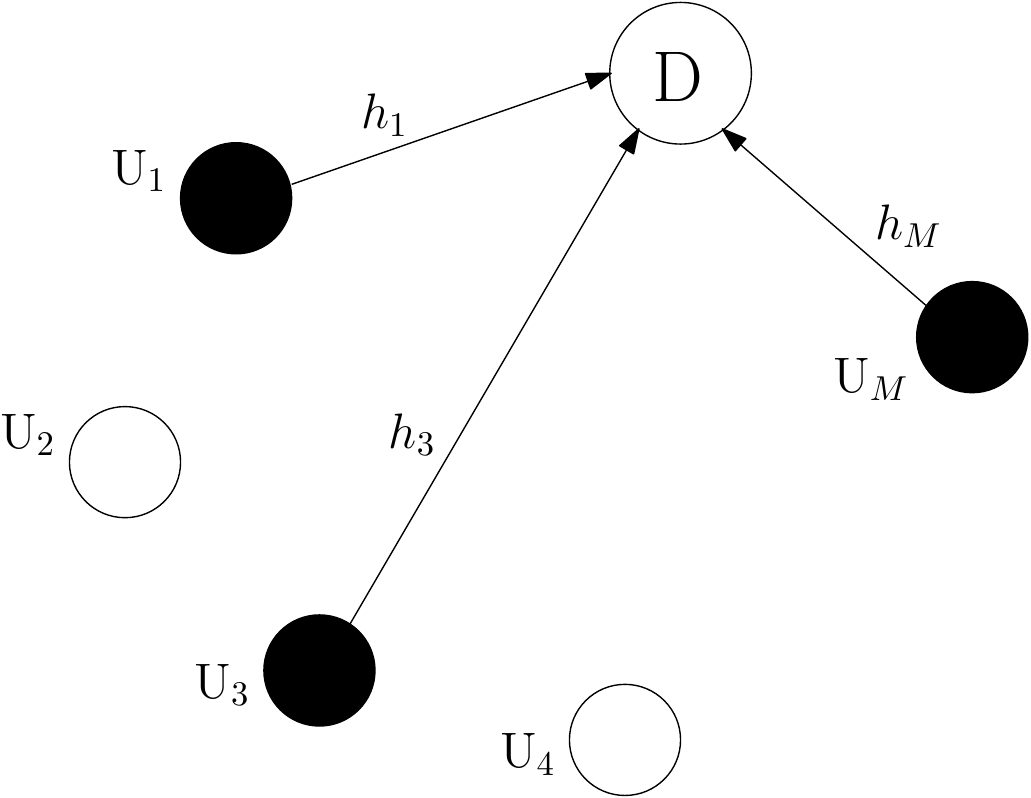}
\caption{M-User multiple access channel. Black circles show active users.}
\label{sysmodel}
\end{figure}

We consider a communication system consisting of $M$ users, U$_1$, U$_2$,..., U$_M$, out of which a subset $\mathcal{S}\ne {\O}$ is active. That is only the users in the set $\mathcal{S}$ are transmitting to a common destination, $D$, as shown in Fig. \ref{sysmodel}. Each active user has $k$ binary information symbols to be delivered at the destination. The transmission model is time slotted, where at each time slot $i\ge 1$, an active user U$_j$ transmits a rateless coded symbol $u_{i,j}$ to the destination. The received signal $y_i$ at the destination at time instant $i$ is given by:
\begin{align}
\textstyle y_i=\sum_{j\in\mathcal{S}}h_ju_{i,j}+n_i,
\end{align}
where $h_j$ denotes the channel coefficient for U$_j$ and $\{n_i\}_{i=1}^{\infty}$ is a sequence of i.i.d. Gaussian random variables with zero mean and unit variance. The set of active users, $\mathcal{S}$, is known to the receiver; however, the transmitters only know whether they belong to $\mathcal{S}$ or not.

From an information theoretic point of view, for a $M$-user Gaussian MAC where there are $M$ users with power constraints $(P_1,P_2,...,P_M)$ and unit variance AWGN noise with zero mean, any rate $R_i$ can be achieved by U$_i$ as follows:
\begin{align}
\label{sumratecap}
\textstyle R_{\mathcal{S}}\triangleq\sum_{i\in\mathcal{S}}R_i<\frac{1}{2}\log_2(1+\sum_{i\in\mathcal{S}}P_i),
\end{align}
where $R_{\mathcal{S}}$ is defined as the sum-rate of active users. Moreover, the upper bound can be achieved when the output signal distribution of each active user follows the Gaussian distribution $\mathcal{N}(0,P_i)$ \cite{EIT}.
\section{Multiple Access Analog Fountain Codes}
Analog fountain codes (AFC) have been recently proposed in \cite{MahyarLetter} as an effective rateless transmission scheme to approach the capacity of wireless channels. AFC is rateless in nature as a potentially limitless number of coded symbols can be generated, which enables the transmitter to effectively adapt to unknown channel conditions. In this section, we first briefly introduce the AFC code and then a general extension of AFC codes, denoted by multiple access analog fountain code (MA-AFC), is proposed.
\subsection{Analog Fountain Codes}
\begin{figure}[!t]
\centering
\includegraphics[scale=0.25]{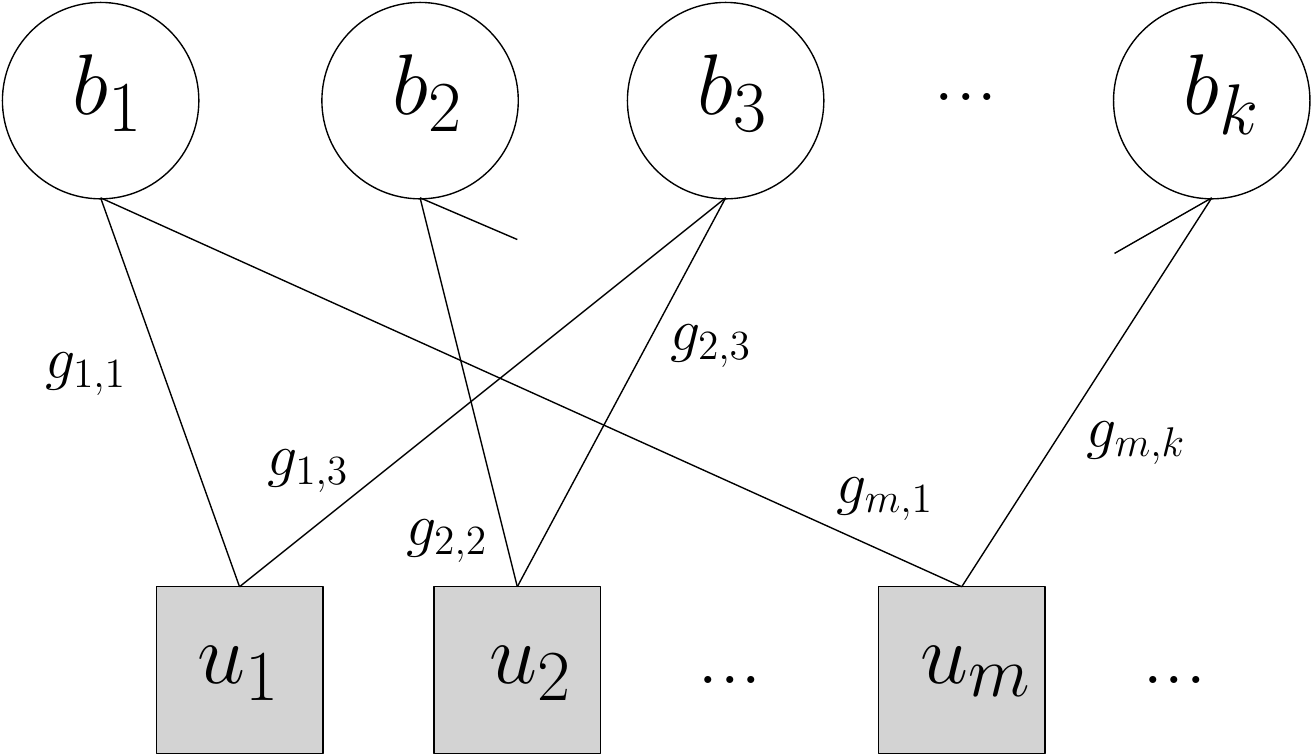}
\caption{Weighted bipartite graph of an AFC with $d_c=2$.}
\label{graph}
\end{figure}

In AFC, the entire message of length $k$ binary symbols is first BPSK modulated to obtain a vector of modulated information symbols, \textbf{b}$_{1\times k}$. To generate each coded symbol, $u_i$, first an integer $d$, called \emph{degree}, is obtained based on a predefined probability distribution function, called \emph{degree distribution}, for $i=1,2,...$ . Then, $d$ randomly selected modulated information symbols are linearly combined with real weight coefficients to generate one coded symbol. For simplicity, we assume that the degree of each coded symbol is fixed, which is denoted by $d_c$ in this paper, and weight coefficients are chosen from a finite \emph{weight set}, $\mathcal{W}_s$, with $f\ge d_c$ positive real members, as follows:
\begin{align}
\mathcal{W}_s=\{w_i\in \mathbb{R}^{+}|i=1,2,...,f\},
\end{align}
where $\mathbb{R}^{+}$ is the set of positive real numbers and $\sigma^2_w\triangleq\frac{1}{f}\sum_{j=1}^{f}w_j^2$ is defined as the average weight set energy. Let $g_{i,j}$ denote the weight coefficient assigned to the $j^{th}$ modulated information symbol in generating the $i^{th}$ coded symbol. It is clear that $g_{i,j}\in\mathcal{W}_S$, if $b_j$ is selected in generating $c_i$; otherwise $g_{i,j}=0$. Let \textbf{G}$_{m\times k}$ denote the matrix of weight coefficients, referred to as the \emph{generator matrix}, where the $j^{th}$ element in the $i^{th}$ row of \textbf{G} is $g_{i,j}$, then AFC encoding process can be shown as a matrix form as follows:
\begin{align}
\label{afcmatrixencoder}
\textbf{u}=\textbf{Gb}',
\end{align}
where \textbf{u} is a $m$ by $1$ vector of coded symbols and $m$ is the number of coded symbols. By considering information and coded symbols as variable and check nodes, respectively, the encoding process of AFC can be further described by a weighted bipartite graph as in Fig. \ref{graph}.

As shown in \cite{MahyarLetter}, if information symbols are selected uniformly at random in the encoding process of an AFC, some information symbols may not have connections to any coded symbol even with a large number of coded symbols, which leads to a decoding error floor. To avoid such an error floor, we have proposed a modified encoder for AFC by maximizing the minimum variable node degree. This way, to generate each coded symbol, $d_c$ information symbols are randomly selected among those which currently have the smallest degrees. The entire data is also precoded using a high rate LDPC code, which has shown to significantly reduce the error floor in AFC codes \cite{MahyarLetter}. Further details of AFC code  design can be found in \cite{MahyarLetter}.

\subsection{Multiple Access Analog Fountain Codes}
Each coded symbol in AFC is a linear combination of modulated information symbols with real weight coefficients. Thus, the sum of two or more AFC coded symbols is still an AFC coded symbol with a larger degree. This motivates us to consider AFCs in a multiple access scenario, where the sum of coded symbols from various users is received at the destination; thus, forming an equivalent AFC code at the destination.

\begin{figure}[!t]
\centering
\includegraphics[scale=0.35]{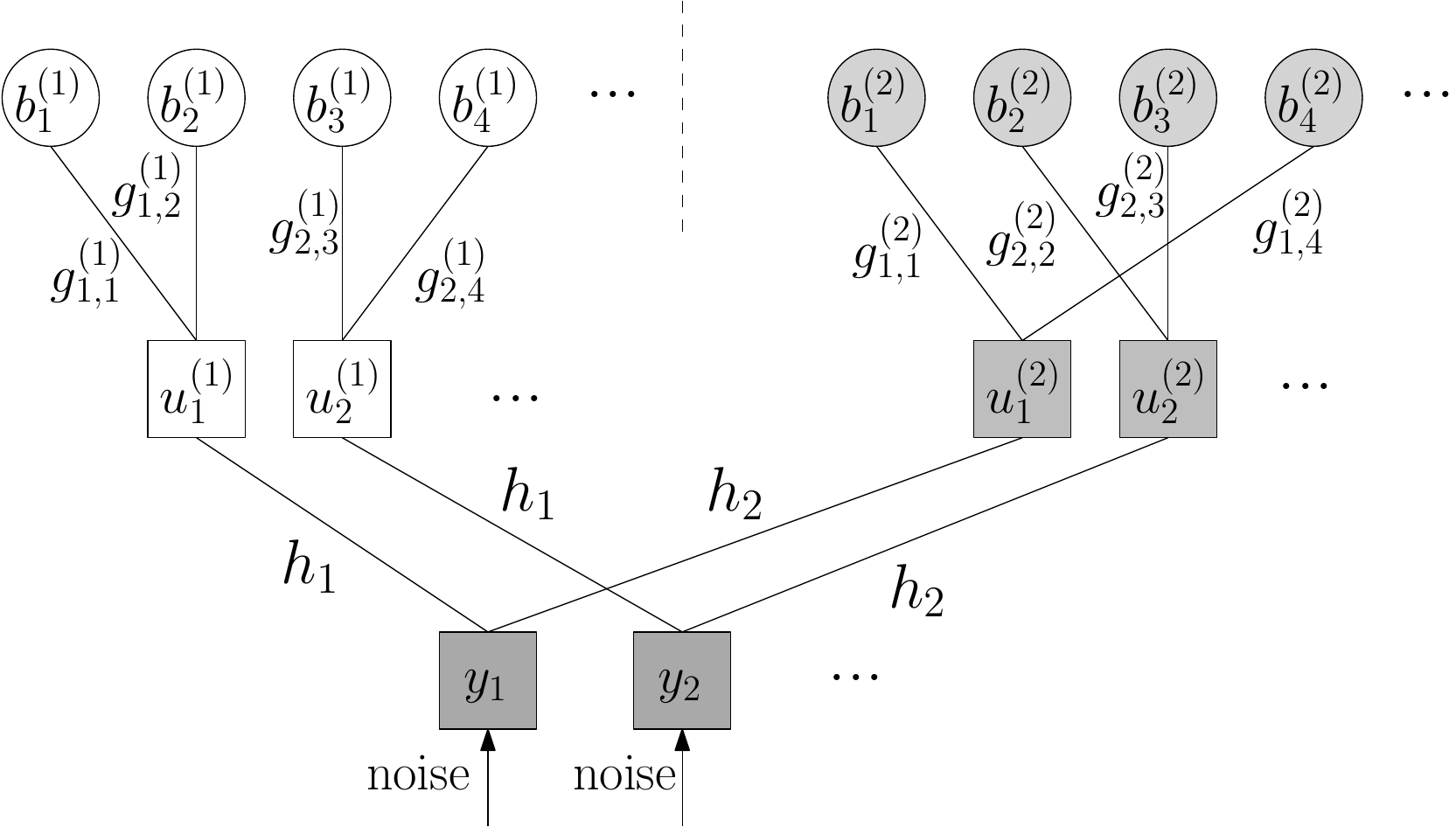}
\caption{Bipartite graph of original AFC code graphs.}
\label{graph211}
\end{figure}

\begin{figure}[!t]
\centering
\includegraphics[scale=0.38]{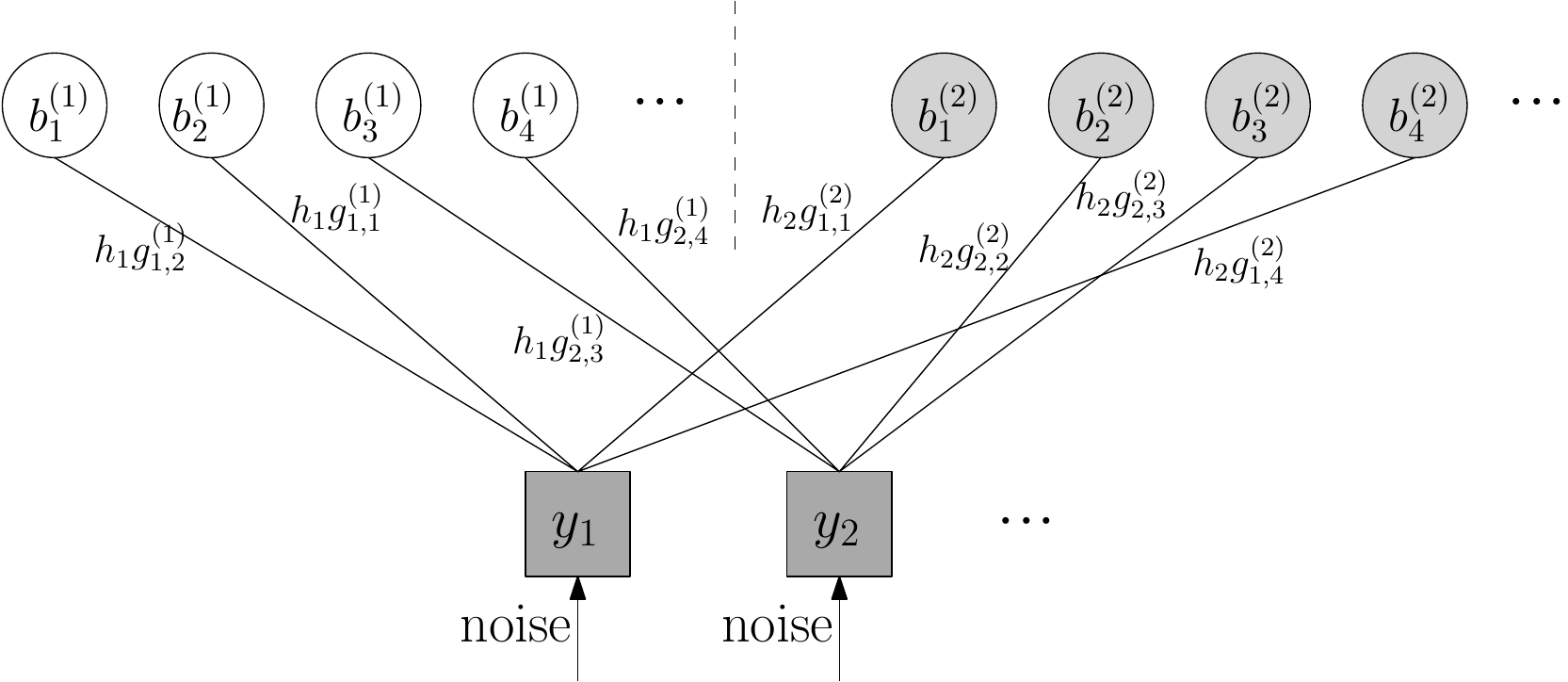}
\caption{Bipartite graph of Equivalent AFC at the destination.}
\label{graph311}
\end{figure}

In the proposed scheme, referred to as multiple access analog fountain code (MA-AFC), each active user $\text{U}_j$ uses an AFC with degree $d_j\ge1$ and a weight set $\mathcal{W}_s$ to generate a potentially limitless number of AFC coded symbols. Let $\textbf{u}^{(j)}_{m\times 1}$ denote the vector of coded symbols generated by U$_j$, and \textbf{G}$^{(j)}_{m\times k}$ denote U$_j$'s generator matrix, where $m$ is the number of coded symbols. Then at time instant $i$, the received signal at the destination, $y_i$, is given by:
\begin{align}
\label{recsig}
\textstyle y_i=\sum_{j\in\mathcal{S}}h_{j}u^{(j)}_{i}+n_i,
\end{align}
According to (\ref{afcmatrixencoder}), we have $u^{(j)}_{i}=\sum_{\ell=1}^{k}g^{(j)}_{i,\ell}b^{(j)}_{\ell}$; thus,  (\ref{recsig}) can be rewritten as follows:
\begin{align}
\label{recsig2}
\textstyle \textbf{y}=\sum_{j\in\mathcal{S}}h_j\textbf{G}^{(j)}\textbf{b}^{(j)'}+\textbf{n},
\end{align}
where $\textbf{b}^{(j)}$ is the information sequence of U$_j$. Eq. (\ref{recsig2}) can be further simplified as follows:
\begin{align}
\label{finalMAAFCmatrix}
\textbf{y}=\textbf{G}\textbf{b}^{'}+\textbf{n},
\end{align}
where $\textbf{G}\triangleq[h_1\textbf{G}^{(1)}|h_2\textbf{G}^{(2)}|...|h_N\textbf{G}^{(N)}]$ and $\textbf{b}\triangleq[\textbf{b}^{(1)}|\textbf{b}^{(2)}|...|\textbf{b}^{(N)}]$. This clearly shows that the received signal at the destination can be seen as a coded symbol of an equivalent AFC, where modulated information symbols are chosen from all active users. It is important to note that the weight coefficients of the equivalent AFC are the weight coefficients of the original AFC codes at each user, multiplied by the gain of the respective channel. More specifically, the equivalent weight set at the destination can be shown by $\mathcal{W}_e\triangleq\bigcup_{i\in\mathcal{S}}\{h_i\mathcal{W}_s\}$ and the code degree $d_e$ of the equivalent AFC code at the destination is $d_e\triangleq\sum_{i\in\mathcal{S}}d_i$. Fig. \ref{graph211} and Fig. \ref{graph311} show the original AFC code at each user and the equivalent bipartite graph of the AFC code at the destination, respectively, when the number of users is 2, i.e., $M=2$.
\subsection{Weight Set Design for MA-AFC}
As mentioned before, to achieve the capacity of the Gaussian MAC, the output signal distribution of each user should follow the Gaussian distribution. In this section, we propose an optimization problem to find the optimum weight set in order to guarantee that AFC coded symbols from each user follow the Gaussian distribution.

Let us consider an AFC with the code degree $d_c$, where information symbols are selected uniformly at random to generate each coded symbol. For a coded symbol $u_i$ and a given real number $c$, we have
\begin{align}
\textstyle p(u_i=c)=p\left(\sum_{\ell\in\mathcal{M}(i)}g_{i,\ell}b_{\ell}=c\right),
\end{align}
where $\mathcal{M}(i)$ is the set of variable nodes connected to the check node $u_i$. Since each information symbol is either -1 or 1 with the same probability of 0.5, and weights are randomly chosen from $\mathcal{W}_s$, then $s_{i,\ell}\triangleq g_{i,\ell}b_{\ell}$ is uniformly distributed as follows:
\begin{align}
p(s_{i,l}=v)=\frac{1}{2f}, ~ |v|\in\mathcal{W}_s.
\end{align}
Furthermore, the mean and variance of $s_{i,\ell}$ are respectively, $m_s=0$ and $\sigma^2_{s}=\frac{1}{f}\sum_{i=1}^{f}(w_i^2)=\sigma^2_w$. Since $s_{i,\ell}$'s are identical and independent random variables, $u_i$ has mean $0$ and variance $d_c\sigma^2_{s}$. Moreover, when $d_c$ is relatively large, according to the central limit theorem, $u_i$ has zero mean Gaussian distribution with variance $d_c\sigma^2_{w}$. However, for a small value of $d_c$, we need to find the optimum weight set in order for the signal distribution to approach the Gaussian distribution. Therefore, we need to find the weight coefficients such that the following condition is satisfied, for a given $\epsilon>0$ and $\delta>0$:
\begin{align}
\label{WeightOptProb}
|p_{\delta}^{(i)}-q_{\delta}^{(i)}|^2\le\epsilon,~~\text{for}~i=1,2,...,i_{max}
\end{align}
where $\textstyle p_{\delta}^{(i)}=p\left((i-1)\delta\le \sum_{j=1}^{d}b_jw_j<i\delta\right)$, $\textstyle q_{\delta}^{(i)}=Q\left((i-1)\delta\right)-Q(i\delta)$, $i_{max}$ is chosen to be large normally larger than 10, and $\textstyle Q(x)=\frac{1}{\sqrt{2\pi}}\int_x^\infty e^{-z^2/2}dz$. It is clear that when condition (\ref{WeightOptProb}) is satisfied, the output signal distribution will be close to the Gaussian distribution with at most $\sqrt{\epsilon}$ difference at $\delta-$distanced signal points. This optimization problem can be numerically solved for different values of $\delta$ and $\epsilon$. For instance, for $f=8$, $\delta=0.2$ and $\epsilon=10^{-4}$, the optimum weight set is as follows: $\{\frac{1}{2},\frac{1}{3},\frac{1}{5},\frac{1}{7},\frac{1}{11},\frac{1}{13},\frac{1}{17},\frac{1}{19}\}$. Since the sum of Gaussian distributed signals is still Gaussian, the distribution of MA-AFC coded symbol is also Gaussian as long as AFC coded symbols of each active user follows a Gaussian distribution.
\section{Decoding of MA-AFC}
Let us consider the bipartite graph of a MA-AFC in Fig. \ref{graph311}. The optimum decoding algorithm for MA-AFC produces the posteriori probability distribution of \textbf{b}, i.e., $P_{\textbf{b}|\textbf{y},\textbf{G}}(.|\textbf{y},\textbf{G})$, which is a sufficient statistic for \textbf{b} in an information theoretic sense, where $\textbf{y}=\textbf{Gb}+\textbf{n}$ as in (\ref{finalMAAFCmatrix}). Here, we present the belief propagation (BP) decoding algorithm for a MA-AFC and then an asymptotic performance analysis of the BP decoder is presented.
\subsection{Belief Propagation Decoding of MA-AFC}
Belief propagation (BP) is an efficient message passing algorithm for estimating the marginal posterior distribution based on the observations. In each iteration of BP, messages are first transmitted from variable to check nodes and then new messages are generated at each check node to be sent back to each of its connected variable nodes.

As variable nodes are binary, then the common choice for the messages is the log-likelihood ratio (LLR) \cite{guo2008multiuser}. Let $L_{r\rightarrow \ell}^{(t)}$ represent the message from variable node $b_r$ to check node $y_{\ell}$ and $L_{\ell\rightarrow r}^{(t)}$ represents the message in the reverse direction at the $t^{th}$ iteration of the BP algorithm. The updating rules for the BP algorithm at the $t^{th}$ iteration ($t\ge1$) can then be shown as follows \cite{guo2008multiuser}:
\begin{align}
\label{update1}
L_{\ell\rightarrow r}^{(t)}&=\text{log}\frac{P\left\{y_{\ell}|b_r=+1,\textbf{G},\{L_{r'\rightarrow \ell}^{(t-1)}\}_{r'\in\mathcal{M}(\ell)\backslash r}\right\}}{P\left\{y_{\ell}|b_r=-1,\textbf{G},\{L_{r'\rightarrow \ell}^{(t-1)}\}_{r'\in\mathcal{M}(\ell)\backslash r}\right\}},\\
L_{r\rightarrow \ell}^{(t)}&=\sum_{\ell'\in\mathcal{N}(r)\backslash  \ell}L_{\ell'\rightarrow r}^{(t)},
\end{align}
where $\mathcal{M}(\ell)\backslash r$ is the set of all variable nodes connected to check node $y_{\ell}$ except variable node $b_r$ and $\mathcal{N}(r)\backslash \ell$ is the set of all check nodes connected to variable node $b_r$ except check node $y_{\ell}$. By considering that the noise variance is 1, we have
\begin{multline}
\nonumber \textstyle P\left\{y_{\ell}|b_r=x,\textbf{G},\{L_{r'\rightarrow \ell}^{(t-1)}\}_{r'\in\mathcal{M}(\ell)\backslash r}\right\}\\
=\sum_{(b_{r'})_{r'\in\mathcal{M}(\ell)}}\frac{1}{\sqrt{2\pi}}e^{-\frac{1}{2}\left(y_{\ell}-\sum_{r'\in\mathcal{M}(\ell)}g_{\ell,r'}b_{r'}\right)^2}\prod_{r'\in\mathcal{M}(\ell)}p_X(b_{r'}),
\end{multline}
where $x\in\{-1,+1\}$ and $p_X(b_r=+1)\propto \exp(L_{r\rightarrow \ell}^{(t)})$. The initial messages $L_{r'\rightarrow \ell}^{(0)}$ correspond to prior LLRs of modulated information symbols, which is $0$ due to the fact that information symbols are equally probable binary random variables. These iterations are repeated until a specific number of iterations are performed or convergence is achieved. After a predefined number of iterations, $T$, the final value of LLRs can be calculated as $L_{r}^{(T)}=\sum_{\ell'\in\mathcal{N}(r)}L_{\ell\rightarrow r}^{(T)}$. A variable node $b_r$ is then decoded as $1$, if $L_{r}^{(T)}>0$; otherwise, it is decoded as $0$.

\subsection{Asymptotic Performance Analysis of MA-AFC based on the BP Decoding}
A commonly used analytical tool for analyzing a BP decoder is density evolution, which calculates the evolutions of message passing in the iterative decoding process. In this paper, we focus on the density evolution analysis in the asymptotic case, when the number of variable and check nodes go to infinity.

Let us refer to information symbols of U$_i$ as Type-X$_i$ variable nodes in the bipartite graph of a MA-AFC code, for $i=1,2,...,N$. The following lemma gives an approximation of LLRs for various types of variable nodes in each iteration of the BP decoder in the asymptotic case, when the number of variable and check nodes are very large.
\newtheorem{lemma}{Lemma}
\begin{lemma}
\label{GeneralDensity}
Let $L^{(t)}_{\text{X}_i\rightarrow \ell}$ denote the message passed from a Type-X$_i$ variable node to a check node $y_{\ell}$ in the $t^{th}$ iteration of the BP decoding algorithm. Then, $L^{(t)}_{\text{X}_i\rightarrow \ell}$ can be approximated by a normal random variable with mean $m^{(t)}_{i}$ and variance $2m^{(t)}_{i}$, where $m^{(t)}_{i}$ can be calculated as follows:
\begin{align}
\label{MAAFCM}
m^{(t)}_{i}=h_i^2\sigma^2_w\frac{d_im}{k}\frac{2}{1+\sigma^2_{Y}},
\end{align}
where $\sigma^2_{Y}=\sum_{j\in\mathcal{S}}h_j^2d_j\sigma^2_w S\left(m^{(t)}_{j}\right)$ and $S(x)=\frac{1}{\sqrt{2\pi}}\int_{-\infty}^{+\infty}\left(1-\tanh(x-y\sqrt{x})\right)e^{-\frac{y^2}{2}}dy$.
\end{lemma}
The proof of this lemma is provided in Appendix \ref{prooflemma}. The BER for each device after $t$ iterations of the BP decoder can then be calculated as $Q\left(\sqrt{m_i^{(t)}}\right)$ \cite{guo2008multiuser}. Let $P_{e,i}$ denote the BER of U$_i$ at the destination, then by using Lemma \ref{GeneralDensity} and for $i,j\in\mathcal{S}$, we have:
\begin{align}
\label{mapprox}
\frac{m_i^{(t)}}{m_j^{(t)}}=\frac{h_i^2d_i}{h_j^2d_j}.
\end{align}
Accordingly, we have:
\begin{align}
\label{pe1pp}
P_{e,i}=Q\left(Q^{-1}(P_{e,j})\sqrt{\frac{h_i^2d_i}{h_j^2d_j}}\right).
\end{align}
This shows that an active user with a higher product of channel gain and code degree, has a lower BER compared to other active users; thus, can be decoded sooner at the destination.
\section{Simulation Results}
\begin{figure}[!t]
\centering
\includegraphics[scale=0.24]{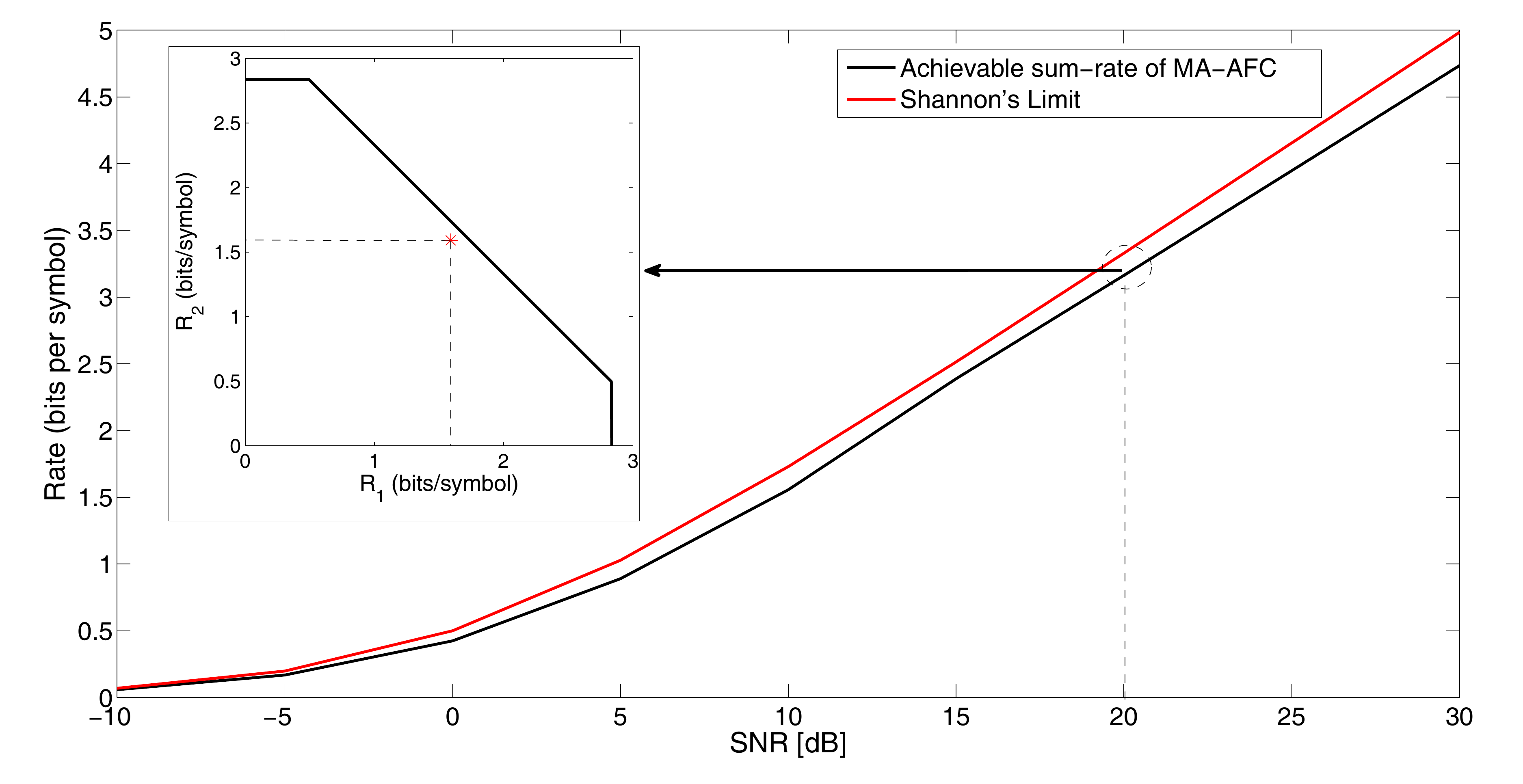}
\caption{Achievable sum-rate of MA-AFC approach versus the received SNR at the destination. The target BER is $10^{-4}$.}
\label{sim1}
\end{figure}

For simulation purposes, we consider $k=1000$ and a AFC code with the weight set $\mathcal{W}_s=\{1/2,1/3,1/5,1/7,1/11,1/13,1/17,1/19\}$. Let us first investigate a 2-user MAC, where the users use AFC with the same weight set. The channel gain and code degree for both users are $h_1=h_2=1$ and $d_1=d_2=4$, respectively. Fig. \ref{sim1} shows the sum-rate achieved by the MA-AFC approach versus the received SNR at the destination at target BER$=10^{-4}$. As can be seen in Fig. \ref{sim1}, MA-AFC can closely approach the sum-rate capacity of the Gaussian MAC in (\ref{sumratecap}), where users are transmitting with the same power. We have also shown the rate achieved by each user at SNR$=20$ dB in Fig. \ref{sim1}, which is very close to the capacity of the Gaussian MAC.

\begin{figure}[!t]
\centering
\includegraphics[scale=0.33]{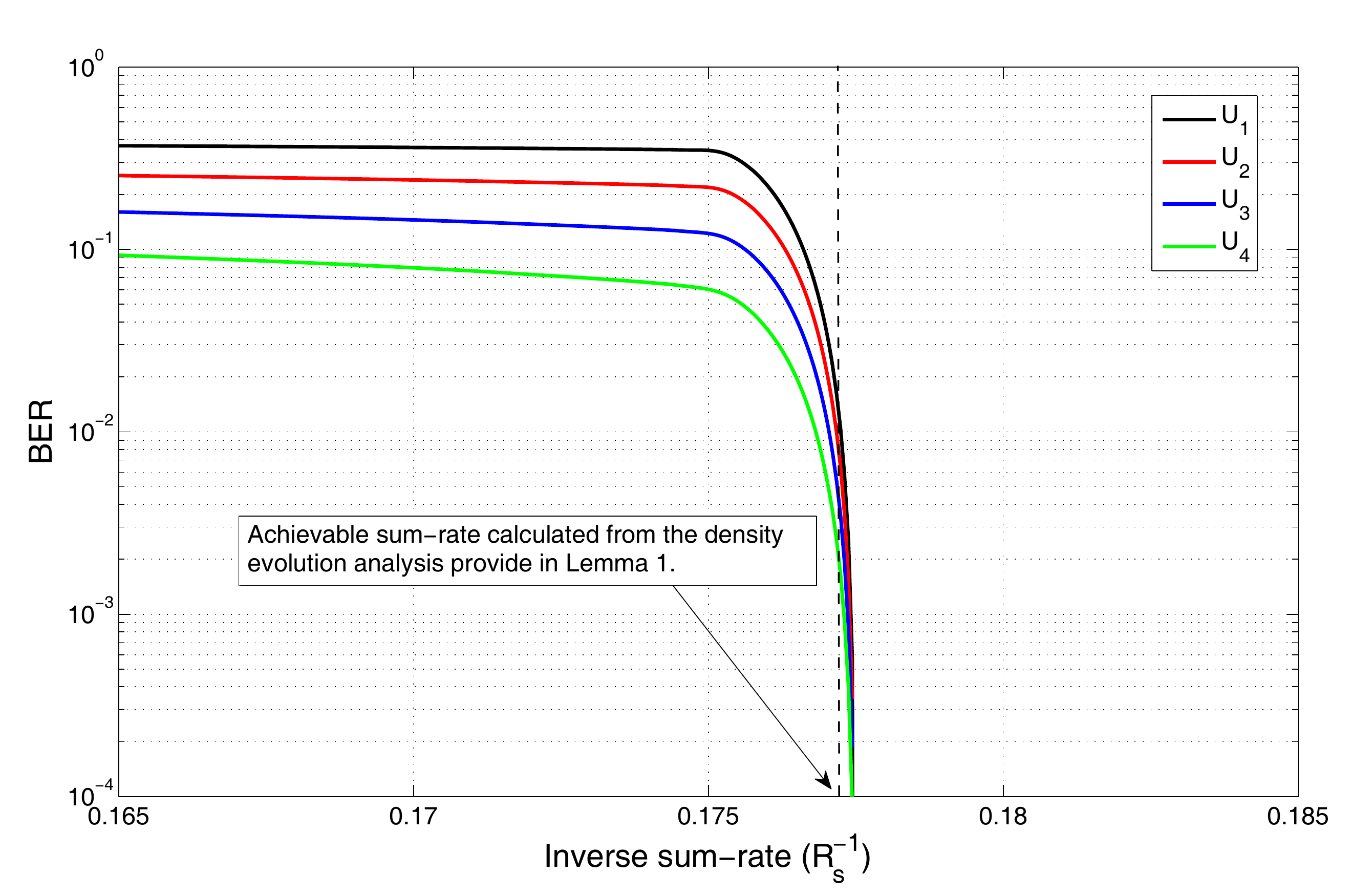}
\caption{BER versus the inverse sum-rate for a 4-user MAC at SNR$=30$ dB. }
\label{sim2}
\end{figure}

Fig. \ref{sim2} shows the bit error rate (BER) versus the inverse sum-rate for a 4-user MAC at $30$ dB received SNR at the destination, where $d_1=d_2=d_3=d_4=4$, and $h_1=1$, $h_2=2$, $h_3=3$, and $h_4=4$. As can be seen in this figure, U$_4$ which has the highest channel gain has a lower BER compared to other users. Furthermore, U$_1$ has the highest BER as it has the lowest product of channel gain and code degree amongst various users. This validates the results of Lemma 1 as the user with the larger of the product of channel gain and code degree will have the lowest BER at the destination. The average BER for each user obtained from the density evolution analysis in Lemma 1 is also shown in Fig. \ref{sim2}, which is very close to the simulation results.

\section{Conclusions}
In this paper, we proposed a novel rateless multiple access scheme based on capacity-approaching analog fountain codes to approach the sum-rate capacity of the multiple access channel in a wide range of SNRs. In the proposed scheme, all users utilize the same AFC code to transmit at the same time and the same channel; thus, forming an equivalent analog fountain code at the destination. The destination then performs joint decoding to recover all users' information symbols. We further analyzed the proposed multiple access analog fountain code by using the density evolution technique. Simulation results show that the proposed approach can approach the sum-rate capacity of the Gaussian multiple access channel in a wide range of SNRs.

\appendices
\section{Proof of Lemma \ref{GeneralDensity}}
\label{prooflemma}
Let us first define $Y_j\triangleq\sum_{r'\in\mathcal{M}(j)\backslash r}g_{j,r'}b_{r'}$ as the $j^{th}$ coded symbol without the $r^{th}$ information symbol. In \cite{guo2008multiuser}, it has been shown that the LLR passed from check node $y_{j}$ to variable node $b_r$ in the $t^{th}$ iteration of the BP algorithm, $L_{j\rightarrow r}^{(t)}$, can be approximated as follows:
\begin{align}
\nonumber L_{j\rightarrow r}^{(t)}=2\frac{f_1(y_j)}{f_0(y_{j})}g_{j,r},
\end{align}
where $f_m(y)$ is defined as follows for $m\in\{0,1\}$:
\begin{multline}
\textstyle \nonumber f_m(y)=\sum_{r'\in\mathcal{M}(j)\backslash r}\frac{1}{\sqrt{2\pi}}e^{-\frac{1}{2}\left(y-Y_j\right)^2}\\
\nonumber \textstyle \times(y-Y_j)^m\prod_{r'\in\mathcal{M}(j)\backslash r}p_X(b_{r'}).
\end{multline}
Furthermore, the mean and variance of $L_{r\rightarrow \ell}^{(t)}$ can be calculated as follows \cite{guo2008multiuser}:
\begin{align}
\label{Mean1L}
\textstyle |\text{E}\left\{L_{r\rightarrow \ell}^{(t)}\right\}|=&2\int_{-\infty}^{+\infty}\frac{f_1^2(y)}{f_0(y)}dy\sum_{\ell'\in\mathcal{N}(r)\backslash \ell}g^2_{\ell',r}\\
\textstyle \text{var}\left\{L_{r\rightarrow \ell}^{(t)}\right\}=&2|\text{E}\left\{L_{r\rightarrow \ell}^{(t)}\right\}|.
\label{Var1L}
\end{align}
Moreover, when the number of variable and check nodes go to infinity, the integral part of (\ref{Mean1L})  tends to $\frac{1}{1+\sigma^2_Y}$ \cite{guo2008multiuser}, where $\sigma^2_Y$ is the variance of $Y_j$. It has been shown that in the asymptotic case, $L_{r\rightarrow \ell}^{(t)}$ is normally distributed with the mean and variance calculated in (\ref{Mean1L}) and (\ref{Var1L}), respectively. More specifically, for a Type-X$_i$ variable node, (\ref{Mean1L}) can be rewritten as follows:
\begin{align}
\textstyle m^{(t)}_i\triangleq|\text{E}\left\{L^{(t)}_{\text{X}_i\rightarrow \ell}\right\}|=\frac{2}{1+\sigma^2_Y}h_i^2\sum_{\ell'\in\mathcal{N}(X_i)\backslash \ell}g^2_{\ell',X_i},
\end{align}
and accordingly $\text{var}\left\{L^{(t)}_{\text{X}_i\rightarrow \ell}\right\}=2m^{(t)}_i$. As we assume that the number of variable and check nodes go to infinity, according to the law of large number, for large code degree $d_i$, $\sum_{\ell'\in\mathcal{N}(X_i)\backslash \ell}g^2_{\ell',X_i}$ can be approximated by $\sigma^2_wd_im/k$, where $m$ is the number of coded symbols. As shown in \cite{guo2008multiuser}, $\sigma^2_Y$ can be calculated as follows:
\begin{align}
\textstyle \nonumber\sigma^2_Y=\sum_{i\in\mathcal{S}}d_ih_i^2\sigma^2_wS(m_i^{(t)}),
\end{align}
where $S(x)=\frac{1}{\sqrt{2\pi}}\int_{-\infty}^{+\infty}\left(1-\tanh(x-y\sqrt{x})\right)e^{-\frac{y^2}{2}}dy$.
This completes the proof.

\bibliographystyle{IEEEtran}
\footnotesize
\bibliography{IEEEabrv,sample2}

\end{document}